\documentclass[%
 reprint,
superscriptaddress,
 amsmath,amssymb,
 aps,
pra,
]{revtex4-1}
\usepackage[utf8]{inputenc}
\usepackage[english]{babel}
\usepackage{aeguill}
\usepackage{ulem}
\usepackage[justification=justified]{caption}
\usepackage{bbold}
\usepackage{mathptmx}
\usepackage{psfrag,graphicx}
\usepackage{dcolumn}
\usepackage{amsmath,amssymb}
\usepackage{bm}
\usepackage{color}
\usepackage{latexsym}
\usepackage{epstopdf}
\usepackage{color}
\usepackage[english]{babel}
\usepackage{latexsym}
\usepackage{psfrag,graphicx}
\usepackage{amsmath}
\usepackage{amssymb}
\usepackage{amsfonts}
\usepackage{bm}
\usepackage{epstopdf}
\usepackage{appendix}
\usepackage{subfig}

\definecolor{mygrey}{gray}{0.35}
\definecolor{myblue}{rgb}{0.2,0.2,0.8}
\definecolor{myzard}{cmyk}{0,0,0.05,0}
\definecolor{mywhite}{rgb}{1,1,1}
\definecolor{mywhite}{rgb}{1,1,1}
\definecolor{myred}{rgb}{1,0.,0.3}

\usepackage[colorlinks=true,citecolor=myblue,linkcolor=myred]{hyperref}

\def\ba{\begin{align}}
\def\enda{\end{align}}
\def\bi{\begin{itemize}}
\def\ei{\end{itemize}}

\def\be{\begin{equation}}
\def\ee{\end{equation}}
\def\bea{\begin{eqnarray}}
\def\eea{\end{eqnarray}}
\def\bse{\begin{subequations}}
\def\ese{\end{subequations}}

\newcommand{\tr}{\mbox{$\mathrm{Tr}$}}
\newcommand{\ket}[1]{|{#1}\rangle}                       % ket
\newcommand{\bra}[1]{\langle {#1}|}                      % bra
\newcommand{\average}[1]{\langle {#1} \rangle}           % media < >

\newcommand{\Ignore}[1]{ }

\def\i{\text{i}}

\DeclareMathOperator{\sech}{sech}

\begin{document}

\preprint{APS/123-QED}

%\title{Exactly solvable time-dependent bidimensional $PT$-symmetric model embedded in $su(1,1)$-dynamical scenario}
\title{Exactly solvable time-dependent pseudo-Hermitian $su$(1,1) Hamiltonian models}

\author{R. Grimaudo}
\address{Dipartimento di Fisica e Chimica dell'Universit\`a di Palermo, Via Archirafi, 36, I-90123 Palermo, Italy}
\address{I.N.F.N., Sezione di Catania, Catania, Italy}

\author{A. S. M. de Castro}
\affiliation{Universidade Estadual de Ponta Grossa, Departamento de F\'{\i}sica, CEP 84030-900, Ponta Grossa, PR, Brazil}

\author{M. Ku\'s}
\address{Center for Theoretical Physics, Polish Academy of Sciences, Aleja Lotnik\'ow 32/46, 02-668 Warszawa, Poland}

\author{A. Messina}
\address{I.N.F.N., Sezione di Catania, Catania, Italy}
\address{Dipartimento di Matematica ed Informatica dell'Universit\`a di Palermo, Via Archirafi, 34, I-90123 Palermo, Italy}

\date{\today}

\begin{abstract}

An exact analytical treatment of the dynamical problem for time-dependent 2x2 pseudo-Hermitian su(1,1) Hamiltonians is reported.
A class of exactly solvable and physically transparent new scenarios are identified within both classical and quantum contexts.
Such a class is spanned by a positive parameter $\nu$ that allows to distinguish two different dynamical regimes.
Our results are usefully employed for exactly solving a classical propagation problem in a guided wave optics scenario.
The usefulness of our procedure in a quantum context is illustrated by defining and investigating the su(1,1) ``Rabi'' scenario bringing to light analogies and differences with the standard su(2) Rabi model.
Our approach, conjugated with the generalized von Neumann equation describing open quantum systems through non-Hermitian Hamiltonians, succeeds in evidencing that the $\nu$-dependent passage from a real to a complex energy spectrum is generally unrelated to the existence of the two dynamical regimes.

%\begin{description}
%\end{description}
\end{abstract}

\pacs{Valid PACS appear here}% PACS, the Physics and Astronomy
\maketitle

%\tableofcontents

\section{Introduction}

The interest towards the study of non-Hermitian Hamiltonians (NHH) has grown exponentially in the last decades and it is still growing.
This is due not only to the applications they have in many different fields of Physics \cite{Moiseyev}, but rather to the relevant role played in better understanding and developing fundamental aspects of Quantum Mechanics.

To appreciate this point it is enough to consider that particular closed systems may often be described by a non-Hermitian Hamiltonian invariant under a space-time inversion (PT-symmetry), implying in turn the idea of a possible extension of Quantum Mechanics \cite{Bender,Mostafazadeh}.
Many decades ago Feshbach employed for the first time non-Hermitian Hamiltonians to represent effectively the coupling between a discrete level and a continuum of states of a given quantum system \cite{Feshbach}.
Such a kind of approach is still largely adopted nowadays to bring to light several worth physical aspects of open quantum systems \cite{Peng}, as for example phase transitions and exceptional points \cite{Rotter}.
An effective non-Hermitian Hamiltonian is characterized by a secular equation with real coefficients, giving rise thus to either real eigenvalues or pairs of complex-conjugated eigenvalues \cite{Sternheim}.
Such a property guarantees that a non-Hermitian Hamiltonian belongs to the class of pseudo-Hermitian operators, provided it is diagonalizable and possesses a discrete spectrum \cite{Mostafazadeh}.
This fact paved the way to significant research on such a kind of specific non-Hermitian Hamiltonians \cite{Scolarici}, whose physical implementations may be found in different contexts, like optical microspiral cavities \cite{Wiersig}, microcavities perturbed by particles \cite{Wiersig1}, or modelling the propagation of light in perturbed medium \cite{Yariv,Boyd}.

In this paper we want to investigate the dynamical problem of a two-level system described by a time-dependent pseudo-Hermitian su(1,1) Hamiltonian.
Such nonautonomous systems were rarely studied in the context pseudo-Hermitian dynamics.
As we show, they may be of  experimental interest, and one of our aims is finding special classes of new exactly solvable cases.
%In this paper we want to investigate the dynamical problem of a two-level system described by a time-dependent \textcolor{magenta}{pseudo-Hermitian} $su(1,1)$ Hamiltonian, with the aim of finding special classes of new exactly solvable cases, possibly of experimental interest.
The reason why we concentrate mainly on the dynamics of a two-level system stems from the fact that the dynamical problem of an $N$-level system characterized by an $su$(1,1) Hamiltonian may be always traced back to that of a two-level system \cite{Ellinas}.
This implies that we may construct the solution of the $N$-level system by knowing that of the related two-level system \cite{Ellinas}.
Furthermore, we know that in conventional quantum mechanics a variety of complicated quantum-mechanical problems can be reduced to a two-level model \cite{Feynman}.
In many contexts, for example nuclear magnetic resonance \cite{Abragham}, quantum information processing \cite{NC} and polarization optics \cite{Born}, essential changes in the system may be described in terms of a two-state dynamics.
%The study of a two-state dynamics plays a fundamental role in many areas of Physics, such as nuclear magnetic resonance \cite{Abragham}, quantum information processing \cite{NC} and polarization optics \cite{Born}.
%Thus, the two-level models studied here should be regarded not as simple toy models but as effective Hamiltonians that can be exploited to investigate the quantum dynamics of more complex physical systems.
The interest towards $su(1,1)$-symmetric dynamical problem finds its reasons in the fact that many physical scenarios exhibit such a kind of symmetry in their Hamiltonian operators.
For example, the dynamics of a $N=2j+1$-level atom in a cascade coupling with a laser beam with time-dependent intensity and in resonance condition (vanishing detuning) is characterized by a time-dependent Hamiltonian embedded in the $su$(1,1) algebra \cite{Ellinas}.
%In this reference the author emphasizes how the solution of the problem for the easiest case of a two-level system enables us constructing the solution of the analogous problem for a generic $N$-level system, as exactly it happens in the case of an su(2) algebra.
Another important $su(1,1)$ physical scenario may be identified in the treatment of squeezed states of the electromagnetic field and scattering of projectiles from simple diatomic molecules \cite{Gilmore}.
These kinds of physical systems, indeed, possess a matrix group structure presenting subdynamics with an $su(1,1)$-symmetry form.
%Moreover, with the help of such a type of \textcolor{magenta}{non-Hermitian} matrices, dissipative phenomena \cite{Peng} may be investigated, for example in quantum optics, and decaying quantum states \cite{Sternheim} may even be described.
%In the last decades the interest towards \textcolor{magenta}{non-Hermitian} matrices has grown since it was firstly demonstrated by Bender and co-workers \cite{Bender} that these matrices could exhibit a real and positive eigenvalue spectrum.
%This fact implies the existence of a lot of new interesting admissible physical scenarios, not well studied before and excluded because of their non-hermiticity.
Moreover, a connection between $PT$-symmetric and $su(1,1)$-symmetric Hamiltonians may be easily found.
The most general 2x2 null-trace matrix representing a Hamiltonian that meets all the conditions of $PT$ quantum mechanics presents indeed the following form \cite{Jones-Smith}

\begin{equation}\label{2x2 PT Matrix}
\begin{pmatrix}
  \alpha & i \beta  \\
i \beta  & - \alpha
\end{pmatrix},
\end{equation}
($\alpha$ and $\beta$ are real) and this class of non-Hermitian matrices is a particular sub-class of the wider class identifying the $su(1,1)$-symmetry matrices.

An important application of $PT$-symmetric Hamiltonian is found in the study and description of the dynamics of the so-called gain and loss systems \cite{Rotter,Croke} which may be encountered and realized in different physical contexts.
These physical systems exhibit several interesting properties.
In particular these systems can present a phase transition related to the $PT$-symmetry breaking \cite{Ruter,Liertzer,Bittner,Schindler,Tripathi}.
In these works, specifically, emphasis is given on how the phase transition may be governed experimentally by manipulating the gain and loss parameters and how it can be justified and related to the fact that in this instance the energy spectrum comes to be complex from real.
One may, therefore, wonder what happens if the parameter(s) governing the reality (and/or complexity) of the spectrum and so the symmetry phase of the Hamiltonian are time-dependent.
%In this paper we wish to investigate the quantum dynamics of time-dependent su(1,1) problems.
From a theoretical point view several efforts are yet necessary for a total comprehension and unifying description of dynamics related to time-dependent non-Hermitian Hamiltonians.
Quite recently, proposals and investigations of fundamental issues have been done \cite{Sergi1,Sergi2,Sergi3,Graefe} and important physical aspects have been brought to light about time-dependent non-Hermitian Hamiltonians \cite{Simeonov, Torosov}.
However, very few attempts are present in literature concerning the identification of classes of exactly solvable scenarios for physical systems described by time-dependent non-Hermitian Hamiltonians.

The paper is organized as follows.
Section \ref{Exact Solutions} is dedicated to the presentation of the mathematical approach to solve the Schr\"odinger equation associated to an su(1,1) Hamiltonian model.
The class of exactly solvable su(1,1) problems is reported in the same section together with the analysis of the corresponding dynamical solutions.
In Sec. \ref{Phys App} the usefulness of our results is illustrated exactly treating a classical and a quantum problem.
Conclusions and remarks are contained in the last subsequent section.

\section{Identification of classes of solvable models and their exact solutions} \label{Exact Solutions}

The group $SU(1,1)$ is not compact and as such it does not have finite-dimensional unitary representations. Its lowest-dimensional faithful matrix representation consists of the set of all $2\times 2$ unit-determinant complex matrices $U$, satisfying the relation
\begin{equation}
\hat{\sigma}^{z} U^{\dagger} \hat{\sigma}^{z} = U^{-1},
\end{equation}
${\hat{\sigma}^x,\hat{\sigma}^y,\hat{\sigma}^z}$ being the standard Pauli matrices.
Generators of this non-unitary representation (i.e. a basis of the corresponding representation of the $su(1,1)$ algebra) can be chosen as
%\begin{equation}
%\hat{K}^{z}= {\hat{\sigma}^{z}\over 2},  \quad \hat{K}^{\pm}=\hat{K}^{y} \pm i\hat{K}^{x},
%\label{GSU11}
%\end{equation}
\begin{equation}
\hat{K}^{0}= {\hat{\sigma}^{z}\over 2}, \quad \hat{K}^{1}= -i{\hat{\sigma}^{y}\over 2}, \quad \hat{K}^{2}= i{\hat{\sigma}^{x}\over 2}.
\label{GSU11}
\end{equation}
%with $\hat{K}^{x}=i\hat{\sigma}^{x}/2$, $\hat{K}^{y} =i\hat{\sigma}^{y}/2$, such that $\hat{\sigma}^{z} \hat{K}^{\dagger} \hat{\sigma}^{z} = \hat{K}$,
They satisfy the relations \cite{Klimov}
%\begin{equation}
%\lbrack \hat{K}^{z},\hat{K}^{\pm}]=\pm \hat{K}^{\pm},\quad\lbrack \hat{K}^{+},\hat{K}^{-}]=-\hat{K}^{z}.
%\end{equation}
\begin{equation}
\lbrack \hat{K}^{1},\hat{K}^{2}]=-i \hat{K}^{0},\quad \lbrack \hat{K}^{1},\hat{K}^{0}]=-i \hat{K}^{2}, \quad \lbrack \hat{K}^{2},\hat{K}^{0}]=i \hat{K}^{1}.
\end{equation}
A $t$-dependent (in general, $t$ is a generic parameter) null-trace $2x2$ $su$(1,1)  matrix is linear combination, with real $t$-dependent coefficients $\omega_0(t), \omega_1(t)$ and $\omega_2(t)$ of the generators $\hat{K}^0$, $\hat{K}^1$ and $\hat{K}^2$, namely
\begin{equation}
H(t)=\omega_0(t)\hat{K}^{0}+\omega_1(t)\hat{K}^{1}+\omega_2(t)\hat{K}^{2}.
\end{equation}
In terms of Pauli matrices it can be cast as
\begin{equation}
\begin{aligned}
H(t)&=\Omega(t)\hat{\sigma}^{z} + i\omega_x(t)\hat{\sigma}^{x} - i\omega_y(t)\hat{\sigma}^{y} \\
&=\Omega(t)\hat{\sigma}^{z}-\omega(t)\hat{\sigma}^{+} + \omega^{\ast}(t)\hat{\sigma}^{-},
\end{aligned}
\end{equation}
where, conventionally, $\hat{\sigma}^{\pm}=(\hat{\sigma}^{x}\pm i \hat{\sigma}^{y})/2$ and $\omega(t)$ is a complex parameter defined by $\omega(t)=\omega_y-i\omega_x\equiv|\omega(t)|e^{i\phi_\omega(t)}$, and $\Omega(t)=\omega_0(t)/2, \omega_x(t)=\omega_2(t)/2, \omega_y(t)=\omega_2(t)/2$.
In this way, in the basis of $\hat{\sigma}^{z}$, the matrix representation of a general non-Hermitian operator $H(t)$ belonging to the $su(1,1)$ algebra results as
\begin{equation}\label{2x2 SU(1,1) Matrix}
H(t)=
\begin{pmatrix}
\Omega(t)    & -\omega(t)\\
\omega^*(t)  & -\Omega(t)
\end{pmatrix}.
\end{equation}
From Eq. \eqref{2x2 PT Matrix} we see that the subclass of $PT$-symmetric $su$(1,1) Hamiltonians is identified by $\phi_\omega=\pi/2$ or equivalently by $\omega_y=0$.

It is important to underline that $su(1,1)$-symmetric Hamiltonians are pseudo-Hermitian, that is, by definition \cite{Mostafazadeh}, there exists at least one non-singular Hermitian matrix $\eta$ such that
\begin{equation}\label{PH rel}
H^\dagger(t)=\eta H(t) \eta^{-1}.
\end{equation}
It is easy to see that the simplest matrix satisfying condition \eqref{PH rel} is
\begin{equation}
\eta=\hat{\sigma}^z=
\begin{pmatrix}
1 & 0 \\
0 & -1
\end{pmatrix}.
\end{equation}
A worth result is that a digonalizable operator is pseudo-Hermitian if and only if its eigenvalues are either real or grouped in complex-conjugated pairs \cite{Mostafazadeh}.
This fact is physically relevant since it turns out to be the feature possessed by the non-Hermitian Hamiltonians resulting by the procedure provided by Feshbach \cite{Feshbach} to describe effectively a quantum system with a discrete spectrum coupled to a continuum.
Pseudo-Hermitian Hamiltonians, thus, result very important in the study of open quantum system \cite{Rotter,Sergi1,Sergi2,Sergi3,Graefe, Graefe}, succeeding in describing particular experimentally detectable physical aspects \cite{Rotter,Ruter,Liertzer,Bittner,Schindler,Tripathi}.

The $t$-parameter-dependent spectrum of $H(t)$ reads $E_\pm(t)=\pm\sqrt{\Omega^2(t)-|\omega(t)|^2}$, hence it is real under the condition $|\omega(t)|^2 < \Omega^2(t)$. The reality of the spectrum is a sufficient and necessary condition for $H(t)$ to be quasi-Hermitian \cite{Mostafazadeh}; the condition of quasi-Hermiticity consists in the existence of a positive-definite matrix $\eta_+$ in the set of the matrices $\eta$ accomplishing the equality in Eq. \eqref{PH rel} \cite{Mostafazadeh}.
It can be verified that such a matrix reads
\begin{equation}
\eta_+=
\begin{pmatrix}
1 & -\omega(t)/\Omega(t) \\
-\omega^*(t)/\Omega(t) & 1
\end{pmatrix}.
\end{equation}
It is positive-definite for $|\omega(t)|^2 < \Omega^2(t)$.
In this case we may identify a new Hilbert space in which $H(t)$ is Hermitian or, in other words, we may define a new scalar product $\average{\cdot| \cdot}_{\eta_+}$ (defining the new Hilbert space), namely $\average{\cdot|\eta_+ \cdot}$ (where $\average{\cdot| \cdot}$ is the standard euclidean scalar product), with respect to which $H(t)$ is Hermitian.
However, if the parameter $t$ represents the time, this condition is not sufficient in order that our Hamiltonian describes a closed quantum physical system.
It can be shown, indeed, that a quasi-Hermitian time-dependent Hamiltonian describes a closed quantum system characterized by a (pseudo-)unitary dynamics only if the positive-definite matrix $\eta_+$ is time-independent \cite{Mostafazadeh1}.
This implies that an $su(1,1)$-symmetry Hamiltonian could describe a closed quantum system only if $\phi_\omega(t)=const.$ and $\Omega(t)$ and $|\omega(t)|$ have the same time-dependence, namely $\omega(t)=|\omega_0| f(t)$ and $\Omega(t)=\Omega_0 f(t)$, with $|\omega_0|^2 < \Omega_0^2$.

For all these reasons, in view of possible dynamical applications of finite dimensional $su(1,1)$-symmetry Hamiltonians $H(t)$ in either classical or quantum contexts, we search solutions of the Cauchy problem ($\hbar=1$)
\begin{equation}\label{Schroed Eq U}
i\dot{U}(t)=H(t) U(t), \quad U(0)=\mathbb{1},
\end{equation}
which for Hermitian Hamiltonians constitutes the time evolution Schr\"odinger equation.
To this end we write the non-unitary operator $U(t)$ in the form
\begin{equation}\label{U Kus}
\begin{aligned}
U(t)&\equiv e^{u_1(t)\hat{\sigma}^+}e^{-u_2(t)\hat{\sigma}^z}e^{u_3(t)\hat{\sigma}^-} \\
&=\begin{pmatrix}
e^{-u_2(t)}+u_1(t)e^{u_2(t)}u_3(t) & u_1(t)e^{u_2(t)} \\
e^{u_2(t)}u_3(t) & e^{u_2(t)}
\end{pmatrix},
\end{aligned}
\end{equation}
getting from Eq. \eqref{Schroed Eq U} the following system of differential equations
\begin{equation}\label{RE}
\left\{
\begin{aligned}
\dot{u}_1(t) &= i\omega(t)-2i\Omega(t)u_1(t)+i\omega^*(t)u_1^2(t), \\
\dot{u}_2(t) &= i\Omega(t)-i\omega^*(t)u_1(t), \\
\dot{u}_3(t) &= -i\omega^*(t)e^{-u_2(t)},
\end{aligned}
\right.
\end{equation}
to be associated with the initial conditions $u_j(0)=0$ ($j=1,2,3$).
Once the first Riccati equation is solved, the remaining two can be simply integrated so that the whole $su(1,1)$-symmetry Hamiltonian problem may be exactly solved.
A similar Riccati equation may be obtained when the analogous problem for the $su$(2) case is trated and an interesting interplay between Physics and Mathematics has recently been reported \cite{MGMN}. 

Since no method is available to solve this Riccati Equation for arbitrary $\Omega(t)$ and $\omega(t)$, then, we look for specific relations of physical interest between the Hamiltonian entries so that the Riccati equation under scrutiny can be solved analytically.
To this end let us consider the following change of variable
\begin{equation}\label{Change of Var}
u_1(t)=i e^{i \phi_\omega(t)} Y(t).
\end{equation}
Plugging this expression into the Riccati equation in Eq. \eqref{RE}, we arrive at the following Riccati-Cauchy problem for the variable $Y(t)$
\begin{eqnarray}\label{RE for Y}
\dot{Y}(t) &=& -|\omega(t)|Y^2(t)-i[2\Omega(t)+\dot{\phi}_\omega(t)]Y(t)+|\omega(t)|, \\
Y(0) &=& 0. \nonumber
\end{eqnarray}
It is quite clear, then, that under the analytical constraint
\begin{equation} \label{Exact Scenarios}
2\Omega(t)+\dot{\phi}_{\omega}(t)=2\nu|\omega(t)|,
\end{equation}
with $\nu$ a time independent real non-negative dimensionless parameter, Eq. \eqref{RE for Y} becomes exactly solvable.

It is possible to adopt a different point of view to examine in depth the meaning of Eq. \eqref{Exact Scenarios}.
To this end, inspired by the seminal paper \cite{Rabi 1954}, we perform the transformation \cite{GdCNM}
\begin{equation}\label{Gen Rabi Transf}
\ket{\psi(t)}=\exp\{i \phi_\omega(t) \hat{\sigma}^z/2\} \ket{\tilde{\psi}(t)},
\end{equation}
getting the following new time-dependent Schr\"odinger equation
\begin{equation}
i\ket{\dot{\tilde{\psi}}(t)}=H_{eff}(t)\ket{\tilde{\psi}(t)},
\end{equation}
with
\begin{equation}
H_{eff}(t)=\left[ \Omega(t)+{\dot{\phi}_\omega(t) \over 2} \right] \hat{\sigma}^z - i |\omega(t)| \hat{\sigma}^y.
\end{equation}
From this expression it is clear why the relation \eqref{Exact Scenarios} is a solvability condition for our problem.
Indeed, the corresponding Schr\"odinger equation
\begin{equation}
i\ket{\dot{\tilde{\psi}}(t)}=|\omega(t)|\left[ 2\nu \hat{\sigma}^z - i \hat{\sigma}^y \right] \ket{\tilde{\psi}(t)},
\end{equation}
may be easily solved, even if the effective Hamiltonian is time-dependent.

The solution $Y_{\nu}(t)$ of the particular Riccati equation, related to a specific value of $\nu$, reads
%\begin{equation}
%Y_{\nu}(t)=|Y_{\nu}(t)|\exp\left[i\varphi_{\nu}(t)\right],
%\end{equation}
%with
\begin{equation}\label{Y-g}
\!\!\! Y_{\nu}(t) = \frac{\sqrt{\nu^{2}-1}\,\tan[\sqrt{\nu^{2}-1}\,\chi(t)]-i\nu\tan^{2}[\sqrt{\nu^{2}-1}\,\chi(t)]}{\nu^{2}\sec^{2}[\sqrt{\nu^{2}-1}\,\chi(t)]-1},
\end{equation}
%\begin{subequations}
%\begin{align}
%|Y_{\nu}(t)| &= \left[ \,\frac{\tan^{2}[\sqrt{\nu^{2}-1}\,\chi(t)]}{\nu^{2}\sec^{2}[\sqrt{\nu^{2}-1}\,\chi(t)]-1}\,\right]^{1/2}, \label{mdY-g} \\
%\varphi_{\nu}(t) &= -\arctan\left[ \,\frac{\nu\tan[\sqrt{\nu^{2}-1}\,\chi(t)]}{\sqrt{\nu^{2}-1}}\,\right], \label{phY-g}
%\end{align}
%\end{subequations}
where the time dependent positive function $\chi(t)$ is defined as
\begin{equation}
\chi(t)=\int_{0}^{t}|\omega(\tau)|d\tau.
\end{equation}

We may identify different classes and related different solutions depending on the value of the parameter $\nu$.
The case $\nu>1$ defines the trigonometric regime with solution $Y_{\nu}^t(t)$ in the form \eqref{Y-g}.
For $0 < \nu < 1$ the solution $Y_{\nu}(t)$ is in the hyperbolic regime having the form
\begin{equation}\label{Y-h}
Y_{\nu}^h(t)= \frac{\sqrt{1-\nu^{2}}\tanh[\sqrt{1-\nu^{2}} \,\chi(t)]-i\nu\tanh^{2}[\sqrt{1-\nu^{2}}\,\chi(t)]}{1-\nu^{2}\sech^{2}[\sqrt{1-\nu^{2}}\,\chi(t)]}.
\end{equation}
%\begin{subequations}
%\begin{align}
%|Y_{\nu}(t)| &= \left[ \,\frac{\tanh^{2}[\sqrt{1-\nu^{2}} \,\chi(t)]}{1-\nu^{2}\sech^{2}[\sqrt{1-\nu^{2}}\,\chi(t)]}\,\right]^{1/2}, \label{mdY-h} \\
%\varphi_{\nu}(t) &= -\arctan\left[  \,\frac{\nu\tanh[\sqrt{1-\nu^{2}}\,\chi(t)]}{\sqrt{1-\nu^{2}}}\,\right]. \label{phY-h}
%\end{align}
%\end{subequations}
The case $\nu=1$ defines the rational regime with
\begin{equation}\label{Y-r}
Y_{\nu}^r(t)= \frac{\chi(t)-i\chi^2(t)}{\chi^{2}(t)+1}.
\end{equation}
%\begin{subequations}
%\begin{align}
%|Y_{\nu}^r(t)| &= \left[ \,\frac{\chi^{2}(t)}{\chi^{2}(t)+1}\,\right]^{1/2},\label{mdY-r} \\
%\varphi_{r}(t) &= -\arctan\left[  \,\chi(t)\,\right].  \label{phY-r}
%\end{align}
%\end{subequations}
Finally, for $\nu=0$ we have the real solution $Y_{0}(t)$ reading
\begin{equation}
Y_{0}(t)=\tanh \left[ \chi(t) \right].
\end{equation}

In this way, through Eq. \eqref{Change of Var}, we may construct the time evolution operator in Eq. \eqref{U Kus} for our exactly solvable scenario of interest.
To this end, it is important to point out that the $SU$(1,1) group elements and then the time evolution operators generated by the Hamiltonians in Eq. \eqref{2x2 SU(1,1) Matrix} depend on only two complex parameters.
Indeed, the Caley-Klein parametrization for the $SU$(1,1) group elements reads
\begin{equation}\label{U Caley-Klein}
U(t)=
\begin{pmatrix}
a(t) & b(t) \\
b^*(t) & a^*(t)
\end{pmatrix},
\end{equation}
with $|a(t)|^2-|b(t)|^2=1$.
Comparing this form with the one given in Eq. \eqref{U Kus} it is easy to derive the following relations
\begin{equation}
u_1={b \over a^*}, \quad u_2=\log(a^*), \quad u_3={b^* \over a^*},
\end{equation}
allowing us to simplify the matrix representation of the time evolution operator in terms of the $u_js$ parameters as follows
\begin{equation}
U(t)=\left(
\begin{array}
[c]{cc}%
\exp[u_{2}^{\ast}(t)] & u_{1}(t)\exp[u_{2}(t)]\\
u_{1}^{\ast}(t)\exp[u_{2}^{\ast}(t)] & \exp[u_{2}(t)]
\end{array}
\right),
\end{equation}
with $e^{u_{2}^{\ast}(t)} e^{u_{2}(t)}(1-|u_1(t)|^2)=1$.
We see that in this case the expressions of the entries are easily readable and symmetric.
Moreover, only two out of the three initial parameters appear.
Then, the evolution operator for our general exactly solvable scenario may be written down as
\begin{widetext}
\begin{equation}
U_{\nu}(t)=\left(
\begin{array}
[c]{cc}
\exp[\mathfrak{r}_{\nu}(t)]\exp[-i\mathfrak{s}_{\nu}(t)] & |Y_{\nu}(t)|\exp[\mathfrak{r}_{\nu}(t)]\exp\left[  i(\mathfrak{s}_{\nu }(t)+\mathfrak{y}_{\nu}(t))\right] \\
|Y_{\nu}(t)|\exp[\mathfrak{r}_{\nu}(t)]\exp\left[ -i(\mathfrak{s}_{\nu }(t)+\mathfrak{y}_{\nu}(t))\right]  & \exp[\mathfrak{r}_{\nu}(t)]\exp[i\mathfrak{s}_{\nu}(t)]
\end{array}
\right),
\label{Usl11}
\end{equation}
\end{widetext}
with
\begin{subequations}
\begin{align}
\mathfrak{r}_{\nu}(t)&=\int_{0}^{t}|\omega(\tau)|\,\text{Re}[Y_{\nu}(\tau)]d\tau, \label{r ni}\\
\mathfrak{s}_{\nu}(t)&=\int_{0}^{t}\Omega(\tau)d\tau+\int_{0}^{t}|\omega(\tau)|\,\text{Im}[Y_{\nu}(\tau)]d\tau, \label{s ni}\\
\mathfrak{y}_{\nu}(t)&=\frac{\pi}{2}+2\nu\int_{0}^{t}|\omega(\tau)|d\tau-2\int_{0}^{t}\Omega(\tau) d\tau+\varphi_{\nu}(t) \label{y ni}, \\
\varphi_\nu(t)&=-\arctan\left[ {\nu\tan[\sqrt{\nu^2-1}\chi(t)] \over \sqrt{\nu^2-1}} \right].
\end{align}
\end{subequations}
%\begin{subequations}
%\begin{align}
%\mathfrak{r}_{\nu}(t)&=\int_{0}^{t}|\omega(\tau)||Y_{\nu}(\tau)|\cos[\varphi_{\nu}(\tau)]d\tau, \label{r ni}\\
%\mathfrak{s}_{\nu}(t)&=\int_{0}^{t}\Omega(\tau)d\tau+\int_{0}^{t}|\omega(\tau)||Y_{\nu}(\tau)|\sin[\varphi_{\nu}(\tau)]d\tau, \label{s ni}\\
%\mathfrak{y}_{\nu}(t)&=\frac{\pi}{2}+2\nu\int_{0}^{t}|\omega(\tau)|d\tau-2\int_{0}^{t}\Omega(\tau) d\tau+\varphi_{\nu}(t) \label{y ni}.
%\end{align}
%\end{subequations}
%Also, note that $\mathfrak{s}_{2\nu}(t)+\mathfrak{y}_{\nu}(t)$ is written as
%\begin{eqnarray}
%\mathfrak{s}_{2\nu}(t)+\mathfrak{y}_{\nu}(t)
%&=& \frac{\pi}{2} + 2\nu\chi_{0}(t)+\varphi_{\nu}(t)-\int_{0}^{t}\Omega(\tau)d\tau\nonumber \\
%&+& \int_{0}^{t}|\omega(\tau)| |Y_{\nu}(\tau)|\sin[\varphi_{\nu}(\tau)]d\tau.
%\end{eqnarray}
Finally, it is easy to verify that the following identity
\begin{equation}
\det [U_{\nu}(t)]=\exp[2\mathfrak{r}_{\nu}(t)]\left( 1-|Y_{\nu}(t)|^2\right) = 1,
\label{id1}
\end{equation}
is fulfilled at any time instant $t$ for arbitrary $\nu$.

\section{Physical Applications} \label{Phys App}

In this section we are going to furnish physically interesting frameworks in which our results may be exploited and could play a relevant role in the solution of the dynamical problems.

\subsection{Physical Implementation in Guided Wave Optics}

We show now how the knowledge of the exact solution \eqref{Usl11} of the dynamical problem may be intriguingly applied to solve a propagation problem in a guided wave optics scenario.
Let us consider two electromagnetic modes counter-propagating in, let us say, the $z$ direction and characterized by the two complex amplitudes $A$ and $B$.
The amplitudes $A$ and $B$ depend on the coordinate $z$ if the two modes propagate in an perturbed medium (e.g. by an electric
field, a sound wave, surface corrugations, etc.), otherwise they are constant.
In the former case, the two amplitudes are mutually coupled in accordance with the following two equations \cite{Yariv}
\begin{equation}\label{Problem electric modes}
\begin{aligned}
{dA(z) \over dz}&=k_{ab}(z)e^{-i\Delta z}B(z), \\
{dB(z) \over dz}&=k_{ba}(z)e^{i\Delta z}A(z),
\end{aligned}
\end{equation}
where $\Delta$ is the phase-mismatch constant and $k_{ab}(z)$ and $k_{ba}(z)$ are complex coupling coefficients determined by the specific physical situation under scrutiny.
Considering the case $k_{ab}(z)=k_{ba}^*(z)=k(z)$ \cite{Yariv}, after few algebraic manipulations, it is possible to verify that the system \eqref{Problem electric modes} may be cast in the following form \cite{Simeonov}
\begin{equation}\label{su(1,1) problem z dependent}
i{dV(z) \over dz}=H(z)V(z),
\end{equation}
where $V(z)=[\tilde{A}(z),\tilde{B}(z)]^T$, with
\begin{equation}
\tilde{A}(z)=A(z)e^{i\Delta z/2}, \quad \tilde{B}(z)=B(z)e^{-i\Delta z/2},
\end{equation}
and
\begin{equation}\label{Hamiltonian z dependent}
H(z)=
\begin{pmatrix}
-\Delta/2 & ik(z) \\
ik^*(z) & \Delta/2
\end{pmatrix}.
\end{equation}

It is worth noticing that $H(z)$ has the same structure of the general 2x2 su(1,1) matrix written in Eq. \eqref{2x2 SU(1,1) Matrix}.
Accordingly, if we write $V(z)=\mathcal{U}(z)V(0)$, then the system turns out in the following Schr\"odinger-Cauchy problem
\begin{equation}\label{U(z) problem}
i{d\mathcal{U}(z)\over dz}=H(z)\mathcal{U}(z), \quad \mathcal{U}(0)=\mathbb{1},
\end{equation}
that is nothing but the problem we studied in the previous section [see Eq. \eqref{Schroed Eq U}] with $t$ replaced by $z$, for which we found sets of exact solutions related to specific relations between the Hamiltonian parameters making the system analytically solvable.
Our class of solvability conditions \eqref{Exact Scenarios}, in this case, reads ($k(z)\equiv|k(z)|e^{i\phi_k(z)}$)
\begin{equation}\label{Solvability Cond z problem}
2\nu|k(z)|+{d\phi_k(z) \over dz}=\Delta.
\end{equation}
It means that, if the space-dependence of $k(z)$ is such that Eq. \eqref{Solvability Cond z problem} is fulfilled for a specific phase-mismatch, then we are able to solve the original system in Eq. \eqref{Problem electric modes}.
More precisely, under the relation \eqref{Solvability Cond z problem}, we are able with our technique to find the operator $\mathcal{U}(z)$ through which we may construct the solutions $A(z)$ and $B(z)$ of the system \eqref{Problem electric modes}, for any initial condition $A(0)$ and $B(0)$.
Thus, Eq. \eqref{Solvability Cond z problem} furnishes special links between $\Delta$ and $k(z)$ turning out in exactly solvable scenarios of two counter-propagating modes in a perturbed medium.
This is just one of the several examples of the so called quantum-optical analogy \cite{Longhi}, namely when space-dependent optical problems may be mapped into time-dependent quantum dynamical ones.

\subsection{Trace and Positivity Preserving Non-Linear Equation of Motion}

%To investigate an SU(1,1) dynamical problem does not immediately imply its applicability to a concrete physical scenario.
The example discussed in the previous subsection provides a problem in the classical optics context where the knowledge of the solutions of the Cauchy problem \eqref{Schroed Eq U}, based on the general $su$(1,1) non-Hermitian Hamiltonian \eqref{2x2 SU(1,1) Matrix}, may be fruitfully exploited to solve Eq. \eqref{U(z) problem}.
In what follows we aim at exploring the applicability of our results in a quantum dynamical context.
We underline that such an objective is not trivial since in the non-Hermitian Hamiltonian-based quantum dynamics conceptual difficulties in the physical interpretation of the mathematical results, may occur.

In the two-dimensional $SU$(1,1) case, differently from $SU$(2), the complex entries $a(t)$ and $b(t)$ appearing in the operator $U(t)$, solution of the Cauchy problem \eqref{Schroed Eq U}, are spoiled of a direct physical meaning.
In the $SU$(2) case, in fact, we may interpret $|a(t)|^2$ and $|b(t)|^2$ as probabilities and then $U(t)$ as the time evolution operator of our quantum dynamical system, while for the $SU$(1,1) case, considered in this paper, $|a|^2 \geq 1$ since $|a(t)|^2-|b(t)|^2=1$ and consequently $U(t)$ cannot be identified as the time evolution generator.
This is intrinsically related to the the dynamics generated by a $su$(1,1) finite-dimensional Hamiltonian.
Indeed, we know that only the infinite dimensional representations of $SU$(1,1) are unitary.
More in general, the crucial problem related to the physical interpretation of the mathematical results we get from the study of the Cauchy problem \eqref{Schroed Eq U} for a generic non-Hermitian Hamiltonian, lies on the fact that the trace of any initial density matrix is not preserved in time.

To recover the necessary normalization condition at any time instant, following the approach introduced in Ref. \cite{Sergi1}, we put
\begin{equation}\label{rescaled}
\rho(t)={\hat{\rho}'(t) \over \text{Tr}\{\hat{\rho}'(t)\}},
\end{equation}
where $\rho'(t)=U(t)\rho'(0)U^\dagger(t)$ and $\dot{U}(t)=-iH(t) U(t)$.

This choice leads to a ``new dynamics'', that is, to a new Liouville-von Neumann equation governing the dynamics of our system, obtained by differentiating Eq.~(\ref{rescaled}), namely
\begin{equation}\label{Eq non lin NHH}
\dot{\rho}(t)=-i[H_0(t),\rho(t)]-\{\Gamma(t),\rho(t)\}+2\rho(t) \text{Tr}\{\rho(t) \Gamma(t)\},
\end{equation}
where we put $H(t)=H_0(t)-i\Gamma(t)$, with $H_0^\dagger(t)=H_0(t)$ and $\Gamma^\dagger(t)=\Gamma(t)$.

From a physical point of view, this equation possesses interesting properties \cite{Graefe} which makes it a valid candidate to describe the quantum dynamics of physical systems characterized by a non-Hermitian Hamiltonian like $PT$-symmetric systems \cite{Schindler, Tripathi}.
The three most important properties to be pointed out are: 1) a pure state remain pure at any time, while the purity of a mixed state, in general, changes in time; 2) the trace and positivity are preserved at any time since the new equation was constructed \textit{ad hoc} to satisfy this condition in order to recover the concept of probability and a statistical interpretation of the quantum dynamics related to non-Hermitian Hamiltonians; 3) the general solution of Eq. \eqref{Eq non lin NHH} reads, of course,
\begin{equation}
\rho(t)={U(t)\rho'(0)U^\dagger(t) \over \text{Tr}\left\{ U(t)\rho'(0)U^\dagger(t) \right\}},
\end{equation}
where $U(t)$ is the (non-unitary) operator satisfying Eq. \eqref{Schroed Eq U}.
Thus the solution of the non-linear problem \eqref{Eq non lin NHH} is traced back to solve our original problem \eqref{Schroed Eq U}.
This circumstance means that, through the procedure exposed in Sec. \ref{Exact Solutions}, we are able to solve the generalized Liouville-von Neumann non-linear equation \eqref{Eq non lin NHH} for the class of time-dependent scenarios identified by the relation \eqref{Exact Scenarios}, whose time evolution operator $U_\nu(t)$ is reported in Eq. \eqref{Usl11}.

Equation \eqref{Eq non lin NHH} was constructed, to some extent, \textit{ad hoc}, merely postulating a way of keeping the proper normalization of the density matrix during the whole evolution. However, one can argue that the form of an evolution equation is dictated by the fact that it should describe a one-parameter positivity  preserving semigroup modeling time evolution of the density matrix of a quantum system. Such demands (semigroup property, positivity-preservation) are natural for a general, ``legitimate'' quantum evolution: the evolution over a given time can be seen as a composition of consecutive evolutions over consecutive, intermediate times and the probability is conserved.
 
Let us consider a general, linear, positivity preserving map,
	\begin{equation}\label{GL}
	\phi_t(\rho)=\mathbb{U}(t)\rho \mathbb{U}^\dagger(t).
	\end{equation}
When $\mathbb{U}(t)$ is a one-parameter semigroup $\mathbb{U}(s+t)=\mathbb{U}(s)\mathbb{U}(t)$, then, of course, $\phi_t(\rho)$ has the semigroup property, i.e. $\phi_s\circ\phi_t=\phi_{s+t}$, which means that the evolution from time $0$ to $s+t$ is composed from the evolution from $0$ to $t$ followed by the evolution from $t$ to $t+s$.
	
Let us now consider the following nonlinear modification of the map $\phi_t(\rho)$,
	\begin{equation}\label{rescaled1}
	\hat{\phi}_t(\rho)=\frac{\phi_t(\rho)}{\tr\left(\phi_t(\rho)\right)}.
	\end{equation}
Using linearity of trace and $\phi_t$ one easily proves that $\hat{\phi}_t$ also has the semigroup property $\hat{\phi}_s\circ\hat{\phi}_t=\hat{\phi}_{s+t}$ \cite{Grabowski}. Since it is clearly positivity- and trace-preserving, it describes a reasonable quantum evolution.
Therefore, Eq.(\ref{rescaled}) itself is less \textit{ad hoc} than it seems and in addition it derives from the same dynamical map generating the von Neumann-Liouville equation when $H=H^\dagger$.

\subsubsection{Quantum dynamics of a su(1,1) ``Rabi'' scenario}

In order to appreciate better the physical aspects of such an equation of motion, we want to study now the ``Rabi'' scenario for the case of $su$(1,1) Hamiltonians and to point out differences and analogies with the $su$(2) case by bringing to light intriguing dynamical aspects.
We know that the `standard' Rabi scenario describes a spin-1/2 subjected to a time-dependent magnetic field precessing around the $\hat{z}$-axis.
The matrix representation of a general $su$(2) Hamiltonian may written as
\begin{equation}
\tilde{H}(t)=\tilde{\Omega}(t)\hat{\sigma}^z+\tilde{\omega}_x(t)\hat{\sigma}^x+\tilde{\omega}_y(t)\hat{\sigma}^y=
\left(
\begin{array}
[c]{cc}%
\tilde{\Omega}(t) & \tilde{\omega}(t)\\
\tilde{\omega}^*(t) & -\tilde{\Omega}(t)
\end{array}
\right),
\end{equation}
with $\tilde{\omega}(t) \equiv \tilde{\omega}_x(t)-i\tilde{\omega}_y(t) \equiv |\tilde{\omega}(t)|e^{i\phi_{\tilde{\omega}}(t)}$ and where $\hat{\sigma}^k$ ($k=x,y,z$) are the Pauli matrices represented in the eigenbasis $\{\ket{\pm}\}$ of $\hat{\sigma}^z$: $\hat{\sigma}^z\ket{\pm}=\pm\ket{\pm}$.
It is easy to see that the consideration of a magnetic field precessing around the $\hat{z}$-axis amounts to consider the three parameters $\tilde{\Omega}$, $|\tilde{\omega}|$ and $\dot{\phi}_{\tilde{\omega}}$ time independent.
Further, the well known Rabi's resonance condition, ensuring a complete periodic population transfer between the two states $\ket{+}$ and $\ket{-}$, acquires the form $\tilde{\Omega}+\dot{\phi}_{\tilde{\omega}}/2=0$.
It is worth to point out that, also when the three parameters are time dependent, the so-called generalized Rabi's resonance condition $\tilde{\Omega}(t)+\dot{\phi}_{\tilde{\omega}}(t)/2=0$ \cite{GdCNM} is a necessary condition to obtain periodic oscillations with maximum amplitude \cite{GdCNM}.

It is possible to convince oneself that the general $su$(1,1) Hamiltonian, whose matrix representation is reported in Eq. \eqref{2x2 SU(1,1) Matrix}, may be written as follows
\begin{equation}\label{Gen NHH}
H(t)=H_0(t)-i\Gamma(t),
\end{equation}
with
\begin{equation}\label{Gen NHH parts}
H_0(t)=\Omega(t) \hat{\sigma}^z, \qquad \Gamma(t)=-\omega_x(t) \hat{\sigma}^x+\omega_y(t) \hat{\sigma}^y,
\end{equation}
and this time we have $\omega(t) \equiv \omega_y(t)-i\omega_x(t) \equiv |\omega(t)|e^{i \dot{\phi}_\omega(t)}$.
We see, then, that we may interpret the $su$(1,1) Hamiltonian as a Rabi problem with a complex transverse magnetic field.
Analogously to the $SU(2)$ case, we may define the Rabi-like scenario for a $SU$(1,1) dynamical problem the case in which the three parameters $\Omega$, $|\omega|$ and $\dot{\phi}_\omega$ are time independent.
Thus, the related solution for the quantum dynamics is given by Eqs. \eqref{Usl11}, \eqref{r ni}, \eqref{s ni} and \eqref{y ni} with $\Omega(\tau)=\Omega_0$, $|\omega(\tau)|=|\omega_0|$ and $\dot{\phi}_\omega(t)=\dot{\phi}_\omega^0$.

\begin{figure}[htp]
%\centering
\begin{center}
\subfloat[][]{\includegraphics[width=0.22\textwidth]{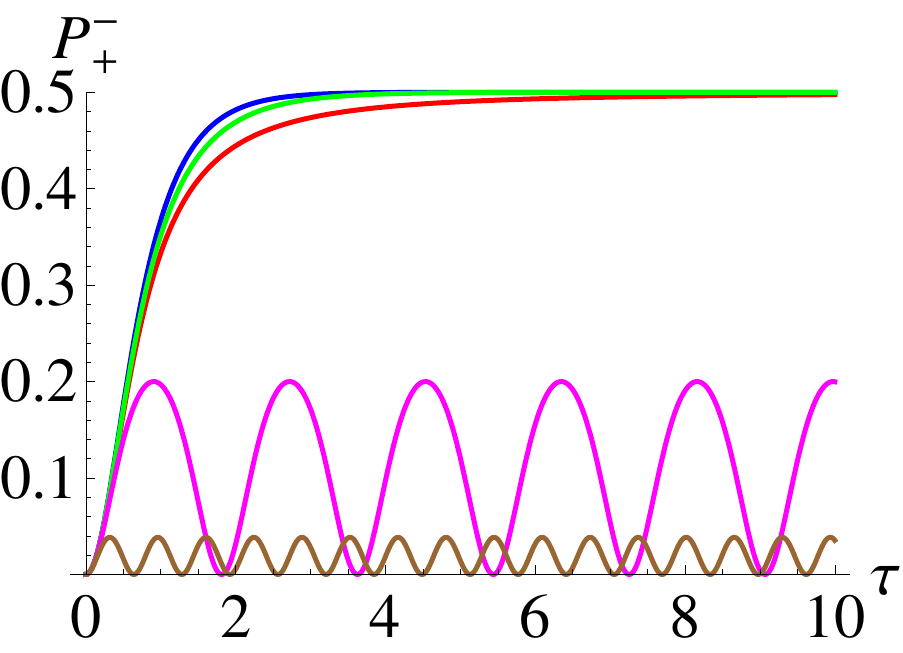}\label{fig:P+-ni}}
\qquad
\subfloat[][]{\includegraphics[width=0.22\textwidth]{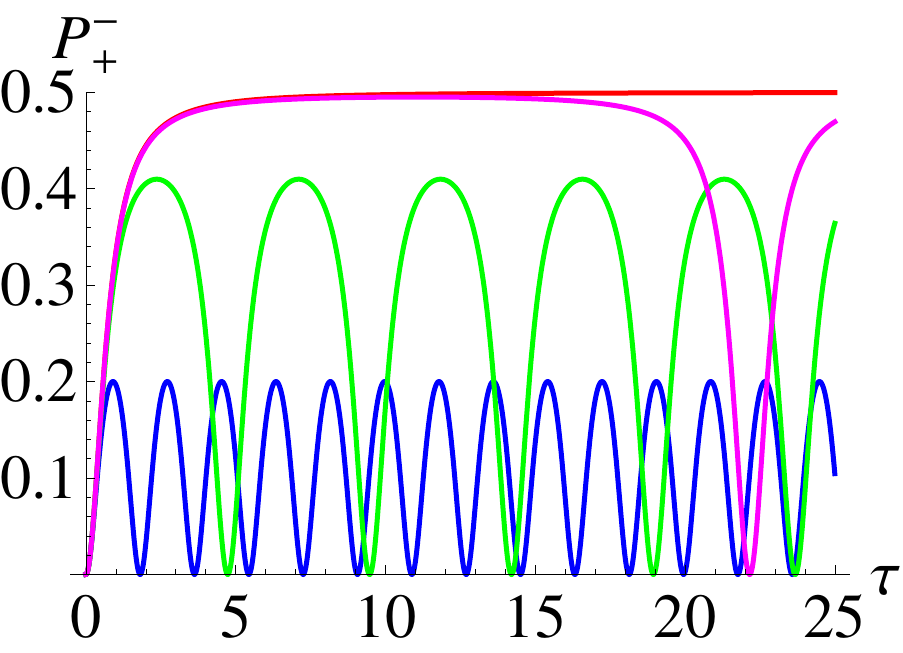}\label{fig:P+-tol}}
\captionsetup{justification=raggedright,format=plain,skip=4pt}%
\caption{(Color online) a) Time dependence of the transition probability $P_+^-$ for different values of $\nu$: $\nu=0;0.7;1;2;5$ correspond to the color blue, green, red, magenta and brown, respectively; b) The plot illustrates ($\nu=2;1.2;1.01;1 \rightarrow$ blue, green, magenta, red) the passage of $P_+^-(t)$ from the oscillatory regime to the plateau regime.}
\end{center}
\end{figure}

We study now the time behaviour of the Rabi's transition probability $P_+^-(t)$, that is the probability to find the system in the state $\ket{-}$ at time $t$ when it is initialized at time $t=0$ in the state $\ket{+}$.
In the framework of the non-linear equation of motion discussed before to describe the quantum dynamics of a system governed by a non-Hermitian Hamiltonian, it means to consider $\rho_0=\ket{+}\bra{+}$.
Considering the non-unitary operator $U(t)$ both in the Caley-Klein \eqref{U Caley-Klein} and our \eqref{Usl11} form, it is easy to see that $P_+^-(t)=\rho_{22}(t)$ results
%If we study, thus, the time behaviour of the Rabi's transition probability $P_+^-(t) \equiv |\average{-|+(t)}|^2$, with $\ket{+(t)}\equiv U(t)\ket{+}$, considering the scenarios identified by Eq. \eqref{Exact Scenarios}, we find
\begin{eqnarray}
%P_+^-(t) = {|b|^2 \over |a|^2+|b|^2} = \frac{\exp[2\mathfrak{r}_{\nu}(t)]}{\exp[2\mathfrak{r}_{\nu}(t)]\left( 1+|Y_{\nu}(t)|^2\right)}. \label{PAST1}
P_+^-(t) = {|b(t)|^2 \over |a(t)|^2+|b(t)|^2} = \frac{|Y_{\nu}(t)|^2}{ 1+|Y_{\nu}(t)|^2}. \label{PAST1}
\end{eqnarray}
In Figs. \ref{fig:P+-ni} and \ref{fig:P+-tol} we report the transition probability $P_+^-$, against the dimensionless time $\tau=|\omega_0|t$, for different values of the parameter $\nu$.
This is done in the case of a Rabi-like scenario which amounts, as explained before, to consider the two parameters $\Omega$ and $|\omega|$, defining the operator $U(t)$ by Eqs. \eqref{Usl11}, \eqref{r ni}, \eqref{s ni} and \eqref{y ni}, independent of time.

We note that we have oscillations when $\nu \geq 1$ of decreasing amplitude and period as long as $\nu$ increases; for $0 \leq \nu < 1$, instead, an asymptotic regime appears.
This constitutes a deep difference between the Rabi scenario in the $SU$(2) and in the $SU$(1,1) case.
In the former, the behaviour of the transition probability $P_+^-(t)$ is always oscillatory in time and different values of $\nu$ are related to different amplitudes of the oscillations.
In the latter, instead, two kinds of time behaviour appear depending on the value of the parameter $\nu$, with 1 as value of separation between the two regimes.
It is important to highlight at this point that the existence of the two regimes, in general, is not related to the reality or complexity of the Hamiltonian spectrum.
The latter, indeed, concerning the ``Rabi'' scenario we are analysing, is $t$-independent, namely $\pm\sqrt{\Omega_0^2-|\omega_0|^2}$, and within the solvability condition \eqref{Exact Scenarios} under scrutiny, it is real if ($\Omega_0 > 0$)
\begin{equation}
\nu > 1+{\dot{\phi}_\omega^0 \over \Omega_0}.
\end{equation}
We see, then, that only if $\dot{\phi}_\omega^0=0$ the $\nu$-dependent transition between the two dynamical regimes coincides with the passage from a real to a complex spectrum.
This happens to be case for the generic $su(1,1)$ 2x2 $PT$-symmetry matrix in Eq. \eqref{2x2 PT Matrix} for which $\phi_\omega(t) = \pi/2$, or for a $t$-independent $su(1,1)$ matrix.
Conversely, if $\dot{\phi}_\omega^0 \neq 0$, two possible interesting cases arise.
Namely, if $\dot{\phi}_\omega^0 < 0$ it means that the transition between the two dynamical regimes ($\nu > 1$ $\rightarrow$ $\nu<1$) occurs while the spectrum keeps its reality, since, in this case, $1+{\dot{\phi}_\omega^0 / \Omega_0} < 1$; on the other hand, if $\dot{\phi}_\omega^0 > 0$ there is a range of values of $\nu$, namely $ 1 < \nu < 1+{\dot{\phi}_\omega^0 / \Omega_0}$, for which the spectrum becomes complex without any appreciable evidence in the dynamical behaviour of the system.

As a last remark we want to highlight that a common feature between the $SU$(2) and the $SU$(1,1) case may be found in the following fact.
It is interesting to note that the Rabi-like resonance condition $\Omega+\dot{\phi}_{\omega}/2=0$ amounts at putting $\nu=0$ and the related curve is the (blue) one in Fig. \ref{fig:P+-ni} being the top limit curve.
We know that in the $SU$(2) case this condition ensures a complete periodic population transfer between the two levels of the system, that is oscillations with maximum amplitude.
Therefore, also in the $SU$(1,1) case, the scenario related to the Rabi's resonance condition is the one with the maximum value for the transition probability at any time.
However, it is important to note that in the $SU$(1,1) case the transition probability, defined according to the framework delineated in Refs. \cite{Sergi1} and \cite{Graefe}, cannot overcome the value of 1/2, meaning that, in this instance, we cannot have complete population transfer.

\section{Conclusions}

The merit of this paper is twofold.
First of all we individuate a non-trivial class of $su(1,1)$ time-dependent Hamiltonian models for which exact solutions of the ``dynamical'' problem: $i\dot{U}(t)=H(t)U(t)$, may be provided.
The direct applicability of our approach to classical optical problems witnesses the usefulness of our method.
Secondly, we construct step by step a reasonable frame within which the knowledge of the non-unitary solution of the above mentioned equation may legitimately exploited as source for generating the time evolution of a generic initial state of the system represented by $H(t)$.
Here ``legitimately'' means that the new dynamical equation for $\rho$ introduced in \cite{Sergi1}, rests on the introduction of a good simple dynamical map generating the standard von Neumann-Liouville equation when the system is described by a Hermitian Hamiltonian.
Exploiting this new point of view, we treat the dynamics of an $su(1,1)$ ``Rabi'' system generating results interpretable within the quantum context.
We analytically evaluate the transition probability $P_+^-(t)$ under the three different regimes highlighted in general terms in Sec. \ref{Exact Solutions}, evidencing remarkable differences from the time behaviour exhibited by the same probabilities in the Rabi su(2) problem.
We have in addition clarified that the passage from a $\nu$-regime to another one is governed by a condition on this parameter which does not coincide with the one ruling the transition from a real (time-independent!) energy spectrum to a complex one.
This result makes evident that such a coincidence might at most be only a particular case ($\dot{\phi}_\omega=0$) in a wider scenario where a direct link between the regime transition and the change in the spectrum of the Hamiltonian does not generally occur.

As a conclusive remark we emphasize that the mathematical analysis and the corresponding results developed in Sec. \ref{Exact Solutions} do not exhaust their potentiality in the quantum context only.
We claim in fact that our method might be of some help in all those situations wherein the behaviour of the system under scrutiny is ruled by a system of non-autonomous first order differential equations exhibiting an $su(1,1)$ intrinsic symmetry.

\section{Acknowledgements}
RG and AM acknowledge stimulating conversation on the subject with A. Sergi.
RG acknowledges for economical support by research funds difc 3100050001d08+, University of Palermo, in memory of Francesca Palumbo.
ASMC acknowledges the Brazilian agency CNPq financial support Grant No. 453835/2014-7.

\end{document}